\documentstyle[twoside,fleqn,npb,epsfig]{article}
\newcommand{\be}{\begin{equation}}
\newcommand{\ee}{\end{equation}}
\newcommand{\bea}{\begin{eqnarray}}
\newcommand{\eea}{\end{eqnarray}}
\newcommand{\nn}{\nonumber}
\newcommand{\AmS}{{\protect\the\textfont2
  A\kern-.1667em\lower.5ex\hbox{M}\kern-.125emS}}

\title{Twist-3 effects for polarized virtual photon structure function 
$g_2^\gamma$}

\author{K. Sasaki \address{Department of Physics, Faculty of Engineering,
        Yokohama National University, \\
        Yokohama 240-8501, Japan}
\thanks{Talk presented at the International Symposium Radcor 2002 
and Loops and Legs 2002, September 8-13, Kloster Banz, Germany.}
       }

\begin{document}

\begin{abstract}
We investigate twist-3 effects in the polarized virtual photon. 
The structure function $g_2^\gamma$, which  exists only for the virtual photon
target and can be measured in future polarized $e^+ e^-$ collider experiments, 
receives both twist-2 and twist-3 contributions. The twist-3 part is 
analyzed in pure QED interaction as well as in LO QCD. We find the  
twist-3 contribution is appreciable for the photon in contrast to the nucleon 
case. 
\end{abstract}

\maketitle

\section{Introduction}

In experiments of polarized deep inelastic lepton-nucleon scattering, we can obtain
information on the two spin-dependent structure functions 
$g_1^{\rm nucl}$ and $g_2^{\rm nucl}$ of the nucleon. In the language of 
operator product expansion (OPE), the twist-2 operators contribute to 
$g_1^{\rm nucl}$ in the leading order of $1/Q^2$. On the other hand, 
$g_2^{\rm nucl}$ receives contributions from both 
twist-2 and twist-3 operators in the leading order. 
The twist-2 part of $g_2^{\rm nucl}$ can be extracted, once $g_1^{\rm nucl}$ is 
measured, by so-called Wandzura-Wilczek (WW) relation \cite{WW}:
\begin{eqnarray}
&&\hspace{-0.5cm} g_{2,tw.2}^{\rm nucl}(x,Q^2)=g_2^{\rm nucl, WW}(x,Q^2)  \nonumber  \\
&&\hspace{0.5cm}\equiv -g_1^{\rm nucl}(x,Q^2)+\int_x^1
\frac{dy}{y}g_1^{\rm nucl}(y,Q^2)~.
\end{eqnarray}
The difference, ${\overline g}_2^{\rm nucl}=
g_2^{\rm nucl}-g_2^{\rm nucl, WW}$, contains the twist-3 contribution.
The experimental data so far obtained show that the twist-3 contribution
to $g_2^{\rm nucl}$ appears to be negligibly small~\cite{E143,E155}.

%\begin{figure}[htb]
%\vspace{9pt}
%\framebox[55mm]{\rule[-21mm]{0mm}{43mm}}
%\caption{Good sharp prints should be used and not (distorted) photocopies.}
%\label{fig:largenenough}
%\end{figure}
%
%\begin{figure}[htb]
%\framebox[55mm]{\rule[-21mm]{0mm}{43mm}}
%\caption{Remember to keep details clear and large enough.}
%\label{fig:toosmall}
%\end{figure}

%\begin{figure}[t]
%\begin{center}
%\vspace{-2cm}
%\includegraphics[height=12cm]{fig1.eps}
%\end{center}
%\vspace{-3cm}
%\caption{The Box-diagram contributions to $g_1^\gamma(x,Q^2,P^2)$ 
%(dashed), $g_2^\gamma(x,Q^2,P^2)$ (solid) and 
%${\bar g}_2^\gamma(x,Q^2,P^2)$ (dash-2dotted) for 
%$Q^2=30$ GeV$^2$ and $P^2=1$ GeV$^2$ for $N_f=3$.}
%\end{figure}

\begin{figure}[htb]
\begin{center}
\includegraphics[scale=0.28]{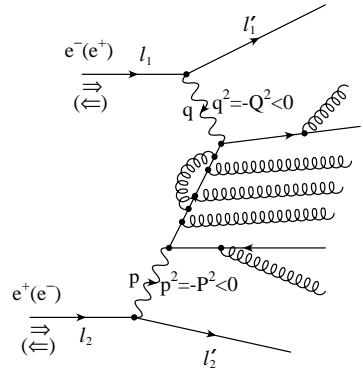}
\caption{Deep inelastic scattering on a polarized virtual photon in 
polarized $e^+e^-$ collision, $e^+e^-\rightarrow {\rm hadrons\  (quarks\  and
\ gluons)}$. The mass squared of the probe (target) photon is $-Q^2$ ($-P^2$) 
($P^2 \ll Q^2$).}
\end{center}
\end{figure}

In recent years, there has been growing interest in the study of spin
structures of photon. The polarized photon structure
functions can be measured by the polarized $e^+ e^-$ collision experiments in the
future linear colliders (Fig.1), where $-Q^2$ ($-P^2$) is the mass squared of
the probe (target) photon.
For the virtual photon target, there appears two structure functions
$g_1^\gamma(x, Q^2, P^2)$ and $g_2^\gamma(x, Q^2, P^2)$, which are the analogues to
the spin-dependent nucleon structure functions $g_1^{\rm nucl}$ and $g_2^{\rm
nucl}$, respectively. 

Now we may ask about the photon structure function
$g_2^{\gamma}$: (i) Does $g_2^{\gamma}$ also receive twist-3 contribution?
(ii) If so, is it small like the nucleon case, or, appreciable?
(iii) Does the WW relation also hold for $g_2^{\gamma}$, in other words,
is the twist-2 part of $g_2^{\gamma}$ expressible in terms of $g_1^{\gamma}$?
(iv) Does any complication occur in the QCD analysis for $g_2^{\gamma}$?
These issues will be discussed~\cite{BSU} in the following.

\section{$g_2^{\gamma}$ in Parton Model}

Let us begin with the analysis of $g_1^\gamma$ and $g_2^\gamma$ in the simple
parton model, in the kinematical region $P^2 \ll Q^2$. 
Evaluating the box diagrams (massless
quark-loops) depicted in Fig.2 with the power corrections  of $P^2/Q^2$ being
neglected, we obtain
\bea
&&\hspace{-0.5cm}g_1^{\gamma({\rm Box})}(x,Q^2,P^2)=
\frac{3\alpha}{\pi}N_f\langle e^4\rangle
\biggl[(2x-1)\ln{\frac{Q^2}{P^2}} \nonumber  \\
&&\hspace{2.5cm} -2(2x-1)(\ln{x}+1)\biggr]~, \label{g1gammaBox} \\
&&\hspace{-0.5cm}g_2^{\gamma({\rm Box})}(x,Q^2,P^2)=
\frac{3\alpha}{\pi}N_f\langle e^4\rangle
\biggl[-(2x-1)\ln{\frac{Q^2}{P^2}}  \nonumber \\
&&\hspace{2.0cm} +2(2x-1)\ln{x}+6x-4\biggr]~,\label{g2gammaBox}
\eea
where 
$x=Q^2/(2p\cdot q)$, $\langle e^4\rangle=
\sum_{i=1}^{N_f}e_i^4/N_f$ with  
$N_f$ being the number of active quark flavours and $\alpha=e^2/4\pi$.

First, note that $g_2^{\gamma({\rm Box})}$ satisfies the 
Burkhardt-Cottingham (BC) sum rule~\cite{BC}, 
\bea
\int_0^1 dx g_2^{\gamma({\rm Box})}(x,Q^2,P^2)=0~.
\eea 
In fact, we will see from the OPE analysis in Sec.3 that the BC sum rule for
$g_2^\gamma$ generally  holds in the deep-inelastic region $Q^2\gg P^2$. Now we
apply the  the WW relation to the above results for $g_1^{\gamma({\rm Box})}$ and
$g_2^{\gamma({\rm Box})}$, and define
\bea
&& \hspace{-0.5cm}g_2^{\gamma{\rm WW(Box)}}(x,Q^2,P^2)\equiv -g_1^{\gamma({\rm
Box})} (x,Q^2,P^2)\nn \\
&&\hspace{1.5cm} +\int_x^1 \frac{dy}{y}g_1^{\gamma({\rm Box})}(y,Q^2,P^2)~.
\eea
The difference, ${\overline g}_2^{\gamma({\rm Box})}=
g_2^{\gamma({\rm Box})}-g_2^{\gamma{\rm WW(Box)}}$, is then given by
\bea
&&\hspace{-0.5cm}{\overline g}_2^{\gamma({\rm Box})}
=\frac{3\alpha}{\pi}N_f\langle e^4\rangle
\,\biggl[(2x-2-\ln{x})\ln{\frac{Q^2}{P^2}} \nn \\
&& -2(2x-1)\ln{x}+2(x-1)+{\ln}^2{x}\biggr]~,\label{g2gammaBoxBar}
\eea
and its $n$-th moment is
\bea
&& \hspace{-0.5cm}{\overline g}_{2,~n}^{\gamma({\rm Box})}=
\frac{3\alpha}{\pi}N_f\langle e^4\rangle \frac{n-1}{n} \nn \\
&& \times \biggl[-\frac{1}{n(n+1)}\ln{\frac{Q^2}{P^2}}
 +\frac{2}{(n+1)^2}-\frac{2}{n^2}\biggr]~. \label{g2gammaBoxBarN}
\eea

\begin{figure}[htb]
\begin{center}
\includegraphics[scale=0.30]{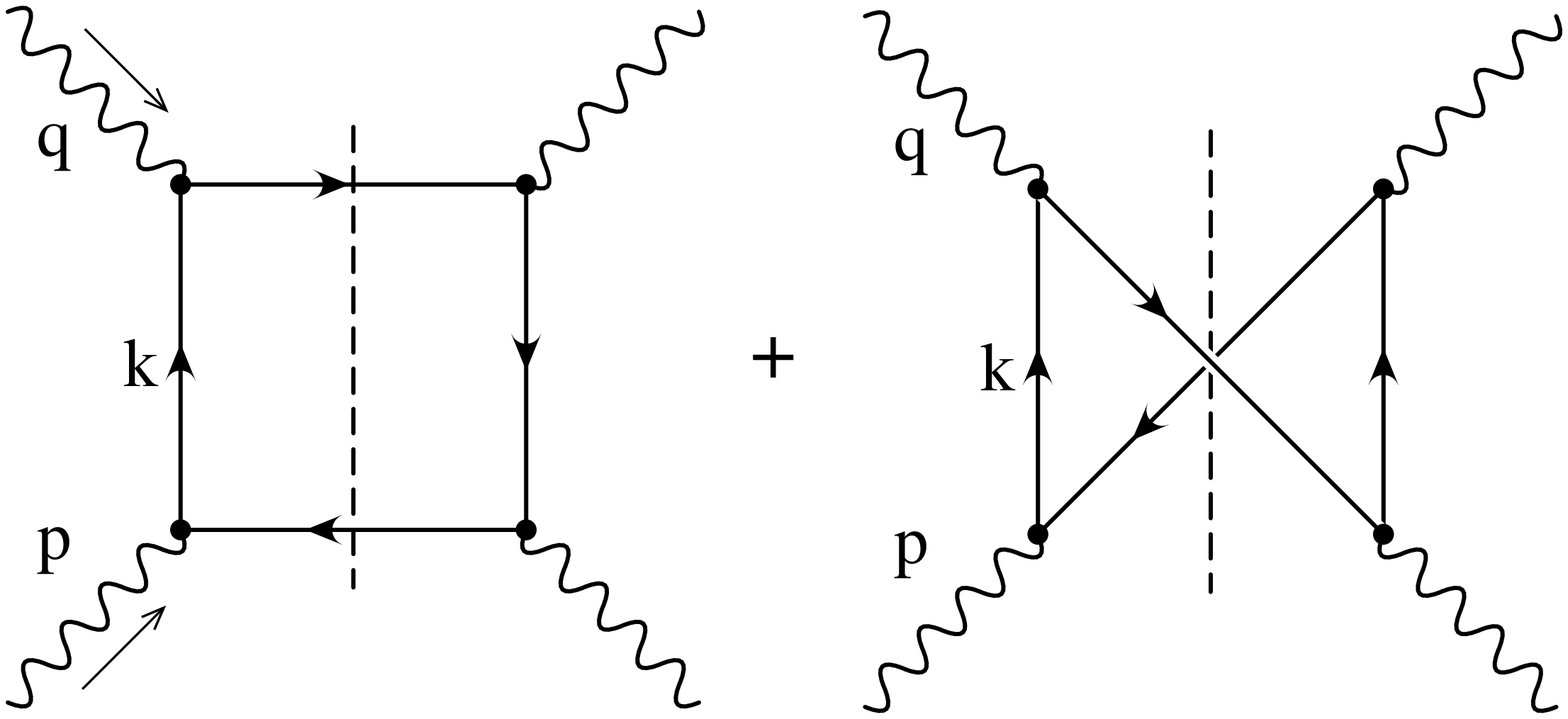}
\caption{The box diagrams contributing to $g_1^\gamma$ and $g_2^\gamma$ in 
pure QED interaction.}
\end{center}
\end{figure}

In Fig.3, we plot the parton model results, $g_1^{\gamma({\rm Box})}$,
$g_2^{\gamma({\rm Box})}$ and ${\overline g}_2^{\gamma({\rm Box})}$
given in  Eqs.(\ref{g1gammaBox},~\ref{g2gammaBox},~\ref{g2gammaBoxBar}),  
as functions of $x$ for $Q^2=30$ GeV$^2$ and $P^2=1$ GeV$^2$.
We can see that ${\overline g}_2^{\gamma({\rm Box})}$ is comparable in
magnitude with $g_2^{\gamma({\rm Box})}$ for large region of $x$.
It is now expected by analogy with the nucleon case  that
${\overline g}_2^{\gamma({\rm Box})}$ arises from the twist-3 effects.
In fact, we  will be convinced in Sec. 3 that {\sl ${\overline
g}_2^{\gamma({\rm Box})}$ is the twist-3 contribution}.

\begin{figure}[htb]
\begin{center}
\vspace{-3.0cm}
\includegraphics[scale=0.40]{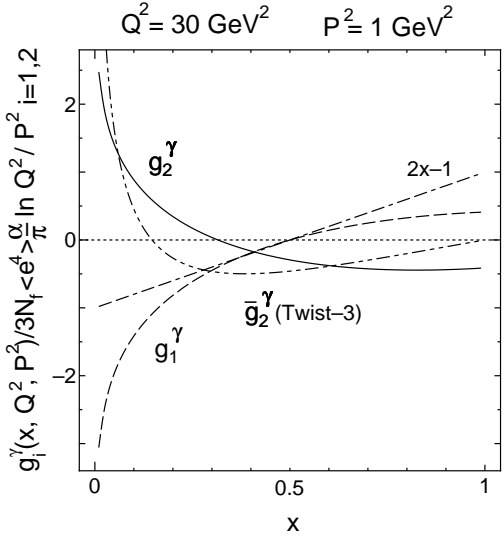}
\vspace{-3.0cm}
\caption{The Box-diagram contributions to $g_1^\gamma$
(dashed line), $g_2^\gamma$ (solid line) and
${\overline g}_2^\gamma$ (dash-2dotted line) for
$Q^2=30$ GeV$^2$ and $P^2=1$ GeV$^2$. The $2x-1$
line shows the leading logarithmic term of $g_1^\gamma$.}
\end{center}
\end{figure}

\section{OPE and Pure QED Effects on $g_2^{\gamma}$}
Applying OPE for the product of two
electromagnetic currents, we get for the $\mu$-$\nu$ antisymmetric part,
\bea
&&\hspace{-0.5cm} i\int d^4x e^{iq\cdot x}T(J_\mu(x)J_\nu(0))^A \nn \\
&&\hspace{-0.5cm}=-i\epsilon_{\mu\nu\lambda\sigma}q^\lambda
\sum_{n=1,3,\cdots}\left(\frac{2}{Q^2}\right)^n
q_{\mu_1}\cdots q_{\mu_{n-1}}  \\
&&\hspace{-0.5cm}\times
\biggl\{
\sum_i E_{(2)i}^n R_{(2)i}^{\sigma\mu_1\cdots\mu_{n-1}}
+\sum_i E_{(3)i}^n R_{(3)i}^{\sigma\mu_1\cdots\mu_{n-1}}
\biggr\},\nn
\eea
where $R^n_{(2)i}$ and $R^n_{(3)i}$ are the twist-2 and twist-3 operators,
respectively, and $E_{(2)i}^n$ and $E_{(3)i}^n$ are corresponding
coefficient functions.
The twist-2 operators $R^n_{(2)i}$ have totally symmetric Lorentz indices
$\sigma\mu_1\cdots\mu_{n-1}$, while the indices of twist-3 operators
$R^n_{(3)i}$  are totally symmetric among $\mu_1\cdots\mu_{n-1}$ but
antisymmetric under $\sigma \leftrightarrow \mu_i$. Thus the
``matrix elements"
of operators $R^n_{(2)i}$ and $R^n_{(3)i}$ sandwiched by two photon states with
momentum $p$ have the following forms:
\bea
&&\hspace{-0.5cm}\langle 0\vert
T(A_{\rho}(-p)R_{(2)i}^{\sigma\mu_1\cdots\mu_{n-1}}A_{\tau}(p))
\vert 0\rangle_{\rm Amp} \nn \\
&&=-ia_{(2)i}^n
{\epsilon_{\rho\tau\alpha}}^{\{ \sigma}p^{\mu_1}\cdots p^{\mu_{n-1}\}}
p^\alpha-({\rm traces}) ~, \nn \label{matTwist2}\\
&&\hspace{-0.5cm}\langle 0\vert
T(A_{\rho}(-p)R_{(3)i}^{\sigma\mu_1\cdots\mu_{n-1}}A_{\tau}(p))
\vert 0\rangle_{\rm Amp}  \nn \\
&&\hspace{-0.5cm}=-ia_{(3)i}^n
{\epsilon_{\rho\tau\alpha}}^{[ \sigma ,~} p^{\{\mu_1~ ]}\cdots
p^{\mu_{n-1}\}} p^\alpha  -({\rm traces}), \nn
\eea
where the suffix \lq Amp\rq\ stands for the amputation of the external
photon lines. 
Then the moment sum rules for $g_1^\gamma$ and $g_2^\gamma$ are written as 
follows:
\bea
&&\hspace{-0.5cm}\int_0^1 dx x^{n-1}g_1^{\gamma}(x,Q^2,P^2)=
\sum_i a_{(2)i}^n E_{(2)i}^n(Q^2), \nn\\
&&\hspace{-0.5cm}\int_0^1 dx x^{n-1}g_2^{\gamma}(x,Q^2,P^2)=\frac{n-1}{n}
 \nn \\
&& \hspace{-0.5cm}\times\biggl[-\sum_i a_{(2)i}^n E_{(2)i}^n(Q^2) 
+\sum_i a_{(3)i}^n E_{(3)i}^n(Q^2)\biggr]. \nn 
\eea

    From this general OPE analysis we conclude: \\
(i)
The BC sum rule holds for $g_2^{\gamma}$,
\be
\int_0^1 dx g_2^{\gamma}(x,Q^2,P^2)=0~.
\ee
(ii) The twist-2 contribution to $g_2^{\gamma}$ is expressed by
the WW relation
\bea
&& \hspace{-0.5cm} -\frac{n-1}{n}\sum_i a_{(2)i}^n E_{(2)i}^n(Q^2)\nn \\
&&\hspace{1.0cm}=\int_0^1 dx x^{n-1}g_2^{\gamma{\rm WW}}(x,Q^2,P^2)~,  
\eea
with
\bea
&& \hspace{-0.5cm}g_2^{\gamma{\rm WW}}(x,Q^2,P^2)\nn \\
&&\hspace{-0.5cm} \equiv -g_1^{\gamma}
(x,Q^2,P^2)+\int_x^1 \frac{dy}{y}g_1^{\gamma}(y,Q^2,P^2)~.
\eea
(iii) The difference, ${\overline g}_2^{\gamma}=g_2^{\gamma}
-g_2^{\gamma{\rm WW}}$, contains  only the twist-3 contribution,
\bea
&& \hspace{-0.5cm}\int_0^1 dx x^{n-1}{\overline g}_2^{\gamma}(x,Q^2,P^2)\nn \\
&& \hspace{1.5cm}=\frac{n-1}{n}\left[ \sum_i a_{(3)i}^n E_{(3)i}^n(Q^2)\right].
\label{Twist3Moment}
\eea

Let us now analyze the twist-3 part of $g_2^{\gamma}$ in pure QED, i.e.,
switching off the quark-gluon coupling,
in the framework of OPE and the renormalization group (RG) method.  In this
case the relevant twist-3 operators are the  quark  and photon operators, which
are given, respectively, by
\bea
&& \hspace{-0.5cm}R_{(3)q}^{\sigma\mu_1\cdots\mu_{n-1}}\nn \\
&&=
i^{n-1} e^2_q~ \overline{\psi}\gamma_5\gamma^{[\sigma,}D^{\{\mu_1]}
\cdots D^{\mu_{n-1}\}}\psi~, \\
&& \hspace{-0.5cm} R_{(3)\gamma}^{\sigma\mu_1\cdots\mu_{n-1}}\nn \\
&&=\frac{1}{4}i^{n-1}
{\epsilon^{[\sigma}}_{\alpha\beta\gamma}F^{\alpha\{\mu_1]}
\partial^{\mu_2}\cdots\partial^{\mu_{n-1}\}}F^{\beta\gamma}~,
\label{photonTwist3}
\eea
where $e_q$ is the quark charge, $D_\mu=\partial_\mu+ieA_\mu$
is the covariant derivative, $F^{\alpha\beta}$ is the photon field strength,  
$\{\ \}$ means complete symmetrization over the indices, while
$[\sigma ,\mu_j ]$
denotes anti-symmetrization on $\sigma\mu_j$, and trace terms are omitted.
With the choice of the above photon operator $R^n_{(3)\gamma}$, we have
$a_{(3)\gamma}^n=1$~.

Solving the RG equation for the coefficient functions 
corresponding to operators $R^n_{(3)q}$ and $R^n_{(3)\gamma}$, we obtain,
to lowest order in $\alpha$, 
\bea
&&\hspace{-0.5cm}E^n_{(3)q}\Bigl(\frac{Q^2}{\mu^2}, \alpha  \Bigr)=
1+{\cal O}(\alpha) \label{Coefficient}
 \\
&&\hspace{-0.5cm}E^n_{(3)\gamma}\Bigl(\frac{Q^2}{\mu^2}, \alpha  \Bigr)
=\frac{\alpha}{8\pi}K^n_{(3)q}\ln{\frac{Q^2}{\mu^2}}+
\frac{\alpha}{4\pi}3e_q^4 B_{(3)\gamma}^n ,\nn 
\eea
where $K^n_{(3)q}$ is the mixing anomalous dimension between the twist-3 photon
operator
$R^n_{(3)\gamma}$ and quark  operator $R^n_{(3)q}$ and is
given by
\be
K^n_{(3)q}=-24e_q^4\frac{1}{n(n+1)}~. \label{anomalousD}
\ee
The ``matrix element" $a^n_{(3)q}$ of the quark operator $R^n_{(3)q}$
between the photon states is calculated to be
\be
a^n_{(3)q}=\frac{\alpha}{4\pi}
\left(-\frac{1}{2}K^n_{(3)q}\ln{\frac{P^2}{\mu^2}}
+3 e_q^4 A^n_{(3)q}\right)~. \label{PhotonMatEle}
\ee
Inserting Eqs.(\ref{Coefficient})-(\ref{PhotonMatEle}) into
(\ref{Twist3Moment}) and remembering  $a_{(3)\gamma}^n=1$, we obtain
for the $n$-th moment of ${\overline g}_2^{\gamma}$ in pure QED,
\bea
&&\hspace{-0.5cm}{\overline g}_2^{\gamma,~ n}\vert_{\rm QED}
=\frac{n-1}{n}\frac{\alpha}{4\pi}3e_q^4
\biggl\{ -\frac{4}{n(n+1)}\ln{\frac{Q^2}{P^2}} \nn \\
&&\hspace{2.5cm} + A^n_{(3)q}+B_{(3)\gamma}^n
\biggr\}~.\label{g2gammaQED}
\eea
The dependence on the renormalization point $\mu$ disappears. And we note
that although $A^n_{(3)q}$ and $B_{(3)\gamma}^n$ are individually
renormalization-scheme  dependent, the sum $A^n_{(3)q}+B_{(3)\gamma}^n$ is
not. The calculation of box diagrams in Fig.2 gives
\be
A^n_{(3)q}+B_{(3)\gamma}^n=8\left\{\frac{1}{(n+1)^2}
-\frac{1}{n^2}  \right\}~.\label{QEDg2gamma}
\ee
Now adding all the quark contributions of active flavours and replacing
$3e_q^4$ in (\ref{g2gammaQED}) with $3N_f\langle e^4\rangle$, we find that
the result is nothing but ${\overline g}_{2,~n}^{\gamma({\rm Box})}$
given in Eq.(\ref{g2gammaBoxBarN}), which is derived from the box-diagram
calculation. Thus it is now clear that ${\overline g}_{2,~n}^{\gamma({\rm
Box})}$ is indeed the twist-3 contribution.
 
\section{QCD Effects on $g_2^{\gamma}$}
We now switch on the quark-gluon coupling and consider the QCD effects on
${\overline g}_2^{\gamma}$, the twist-3 part of $g_{2}^{\gamma}$. In the
nucleon case, the analysis of ${\overline g}_2^{\rm nucl}$,
the twist-3 part
of the structure function $g_2^{\rm nucl}$, turns out to be very
complicated~\cite{ShuVain}.
This is due to the fact that the number of participating
twist-3 operators  grows with spin (moment of ${\overline g}_2^{\rm nucl}$) and
that  these operators mix among themselves through renormalization.  Therefore,
the $Q^2$ evolution equation for the moments of ${\overline g}_2^{\rm nucl}$
cannot be written in a simple form, but in a sum of terms,  the number of which
increases with spin. 

The same is true for ${\overline g}_2^{\gamma}$. 
However, in certain limits the analysis for the moments of  
${\overline g}_2^{\gamma}$ becomes tractable. One is when $n$ is a small number   
and the other is the large $N_C$ (the number of colours) limit for the analysis 
of ${\overline g}_2^{\gamma (NS)}$,
the flavour nonsinglet part of ${\overline g}_2^{\gamma}$. Indeed, for
$n=3$ (the non-trivial lowest moment), we can get all the information on the
necessary anomalous dimensions of participating  operators, and thus we obtain the
LO QCD prediction for the third moment of ${\overline g}_2^{\gamma}$~\cite{BSU}. 
On the other hand, for large $N_C$,  we can evade the
problem of operator mixing for ${\overline g}_2^{\gamma (NS)}$, and obtain the
moments of ${\overline g}_2^{\gamma (NS)}$ in a compact form for all $n$.

In the case of ${\overline g}_2^{{\rm nucl}(NS)}$, 
the twist-3 and flavour nonsinglet part of the  nucleon structure
function $g_2^{\rm nucl}$, it has been observed~\cite{ABH,KS3}
that at large $N_C$, the operators involving gluon field strength
$G_{\mu\nu}$ decouple from the evolution equation of ${\overline g}_2^{{\rm
nucl}(NS)}$, and the whole contribution in LO is represented by one type of
operators. In the photon case, the relevant twist-3 operators for 
${\overline g}_2^{\gamma (NS)}$ are
\bea
&& \hspace{-0.5cm}R_{(3)F}^{\sigma\mu_1\cdots\mu_{n-1}}\nn \\
&&=
i^{n-1}~ \overline{\psi}\gamma_5\gamma^{[\sigma,}D^{\{\mu_1]}
\cdots D^{\mu_{n-1}\}}Q^{\rm ch}\psi~, 
\eea
and the photon operators $R_{(3)\gamma}^n$ given in
Eq.(\ref{photonTwist3}). Here $D_\mu=\partial_\mu-ig A_\mu^aT^a+ieA_\mu$ is the 
covariant derivative, and
$Q^{\rm ch}$  is the quark-charge factor and defined by $Q^{\rm ch}=Q^2-\langle e^2
\rangle  {\bf 1}$, where $Q$ is the $N_f \times N_f$ quark-charge matrix,
$\langle e^2\rangle= \sum_{i=1}^{N_f}e_i^2/N_f$ and
${\bf 1}$ is an $N_f \times N_f$ unit matrix. 
In the approximation of neglecting terms of order {$\cal O$}($1/{N_C^2}$) 
and thus putting $2C_F=C_G$,
the mixing anomalous dimensions between $R_{(3)F}^n$ and other 
hadronic (quark and gluon) operators turn out to vanish. Those which remain 
non-zero are only the $(F,F)$ element and the mixing anomalous dimension between 
$R_{(3)F}^n$ and the photon operator $R_{(3)\gamma}^n$: 
\bea
{\hat \gamma}_{n,FF}^{(0)}&=&8C_F(S_n-\frac{1}{4}-\frac{1}{2n})~, \\
&& {\rm with}\quad S_n=\sum_{j=1}^n\frac{1}{j}~,\nn \\
 K_{n,F}^{(0)}&=&-24N_f(\langle e^4\rangle-\langle
e^2\rangle^2)\frac{1}{n(n+1)}~.
\label{anomalousLargeNC}
\eea
The corrections are of ${\cal O}(1/N_C^2)$, about
$10 \%$ for QCD ($N_C=3$).
Using the above results, we find that, for large $N_C$~, the $n$-th moment of
${\overline g}_2^{\gamma(NS)}$  in LO QCD is given by
\bea
&&\hspace{-0.5cm} \int_0^1 dx x^{n-1}{\overline g}_2^{\gamma(NS)}(x,Q^2,P^2)
\nn \\
&&=\frac{n-1}{n}~\frac{\alpha}{4\pi}~\frac{2\pi}{\beta_0\alpha_s(Q^2)}
K_{n,F}^{(0)}
\frac{1}{1+{\hat \gamma}_{n,FF}^{(0)}/2\beta_0}\nonumber\\
&&\qquad\times   \biggl\{
1-\left(\frac{\alpha_s(Q^2)}{\alpha_s(P^2)}\right)^{
{\hat \gamma}_{n,FF}^{(0)}/2\beta_0+1}\biggr\}~, \label{LargeNCformula}
\eea
where $\alpha_s(Q^2)$ is the QCD running coupling constant and 
$\beta_0=(11N_C-2N_f)/3$ is the one-loop coefficient of the $\beta$ function.

We perform the Mellin transform of Eq.(\ref{LargeNCformula})
to get ${\overline g}_2^{\gamma(NS)}(x,Q^2,P^2)$ as a function of $x$.
The result is plotted in Fig.4.  Comparing with the pure QED box-graph
contribution, we find that the LO QCD effects are
sizable and tend to suppress the  structure
function ${\overline g}_2^{\gamma(NS)}$ both in the large $x$ and small $x$
regions, so that the vanishing $n=1$ moment of ${\overline g}_2^{\gamma(NS)}$,
i.e. the BC sum rule, is preserved.

\begin{figure}[htb]
\begin{center}
\vspace{-3.0cm}
\includegraphics[scale=0.40]{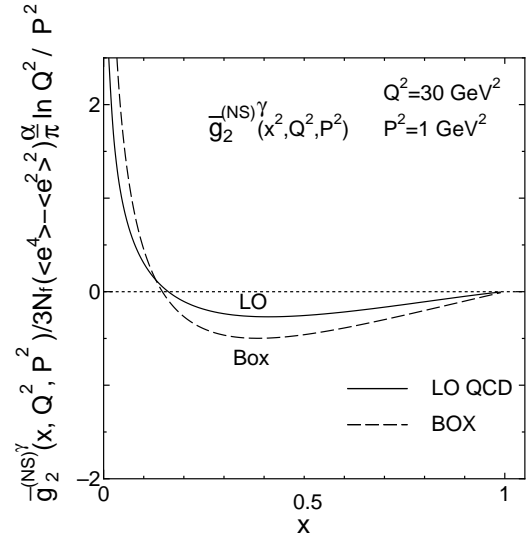}
\vspace{-3.5cm}
\caption{The Box-diagram (dashed line) and the QCD LO (solid line)
contributions for large $N_C$
to the flavour nonsinglet photon structure function
${\overline g}_2^{\gamma(NS)}(x,Q^2,P^2)$
for $Q^2=30$ GeV$^2$ and $P^2=1$ GeV$^2$ for $N_f=3$.}
\end{center}
\end{figure}

\section{Conclusion}

We have analyzed the twist-3 effects
in $g_2^\gamma$ for the virtual photon target, in pure QED
interaction as well as in LO QCD. We have found that the twist-3
contribution is appreciable for the photon in contrast to the nucleon
case. In this sense, the virtual photon structure function $g_2^\gamma$
provides us
with  a good testing ground for studying the twist-3 effects.
We expect that the future polarized version of 
$e^+e^-$ colliders may bring us important information on spin structures 
of photon.

\vspace{0.5cm}
K.S.  thanks H. Baba, J. Soffer and T. Uematsu for useful discussions.
%%%%%%%%%%%%%%%%%%%%%%%%%%%%%%%%%%%%%%%%%%%%%%%%%%%%%%%%%%%%

\end{document}